\documentclass{aa}
\usepackage{txfonts}
\usepackage{psfig}
\usepackage{natbib}

\bibpunct{(}{)}{;}{a}{}{,}

\begin{document}

\def\eps{\varepsilon}
\def\aap{A\&A}
\def\apj{ApJ}
\def\apjl{ApJL}
\def\mnras{MNRAS}
\newcommand\schw{Schwarzschild\ }
\def\aj{AJ}
\def\nat{Nature}
\def\aaps{A\&A Supp.}
\def\degs{$^\circ $}
\def\msun{{\,M_\odot}}
\def\rsun{{\,R_\odot}}
\def\lsun{{\,L_\odot}}
\def\sgra{Sgr~A$^*$}
\def\medd{\dot{M}_{\rm Edd}}
\def\ledd{{L}_{\rm Edd}}
\newcommand\mdot{\dot{m}}
\def\simlt{\lower.5ex\hbox{$\; \buildrel < \over \sim \;$}}
\def\simgt{\lower.5ex\hbox{$\; \buildrel > \over \sim \;$}}

\def\del#1{{}}
\def\C#1{#1}

\title{Using close stars as probes of hot accretion flow in \sgra}
\titlerunning{Close stars and hot accretion flow in \sgra}
\author{Sergei Nayakshin} \authorrunning{S. Nayakshin}
\institute{Max-Planck-Institut f\"{u}r Astrophysik,
Karl-Schwarzschild-Str.1, D-85741 Garching, Germany } \date{\today}

\abstract{The extremely hot and tenuous accretion flow in the
immediate vicinity of \sgra\ is believed to be invisible (too dim) in
the X-ray band, except for short X-ray flares. Here we point out that
during pericenter passages, close brightest stars irradiate the inner
region of the accretion flow, providing a plenty of optical/UV
photons. These seed photons are Compton up-scattered by the hot
electrons of the accretion flow to higher frequencies, some into the
X-ray band, potentially making the innermost accretion flow much
brighter in X-rays than usual. We propose to use coordinated near
infra-red and X-ray observations of close star passages to put
constraints onto \sgra\ accretion theories.  The absence of a
noticeable change in the steady emission of \sgra\ as observed by {\em
Chandra} in the year 2002, when the star named S2 passed through a
pericenter of its orbit, already rules out the hotter of the
``standard'' Advection-Dominated Accretion Flows. The less dense
accretion flows, in particular the model of \cite{Yuan03}, passes the
test and is constrained to accretion rates no larger than
$\sim$~few$\times 10^{-7} \msun$~year$^{-1}$.  } \maketitle

\section{Introduction}

\sgra\ is the super massive black hole (SMBH) in the center of our
Galaxy \citep{Reid99, Schoedel02, Ghez03a}. The gas from the winds of
the surrounding young massive stars should be able to maintain the
accretion rate on the SMBH at a sizable fraction of the Bondi
accretion rate {\em if the SMBH can accept the gas at this rate}. The
Bondi accretion rate is estimated to be $\sim 10^{-6} \msun$
year$^{-1}$ \citep{Baganoff03a}, whereas the linear polarization
measurements at a range of radio wavelengths constrain the accretion
rate onto the SMBH to a value of $\sim 10^{-8}-10^{-7}
\msun$~year$^{-1}$ \citep[e.g.,][]{Bower03}. The latter estimate is
based on somewhat uncertain assumptions about the magnetic field
equipartition and absence of many magnetic field reversals
\citep{Ruszkowski02}. 

It would certainly be beneficial to have another physically
independent method to constrain the properties of the accretion flow
near the event horizon. Unfortunately, in the X-ray band, it appears
that the {\em quiescent} \sgra\ emission is resolved
\citep{Baganoff03a} and can be explained by thermal bremsstrahlung
from the capture radius region \citep{Quataert02}. The inner flow may
thus be simply invisible to the observer unless there is a sizable
X-ray emission from a jet \citep[e.g.,][]{Yuan02}, or during strong
X-ray flares \citep{Baganoff01}.

In this Letter we note that comptonization of the radiation of the
closest of the observed bright stars by the hot accretion flow may be
used to this effect. These stars produce as much as $L_*\sim 10^5
L_\odot \sim 4 \times 10^{38}$ erg/sec in optical-UV radiation, the
amount far greater than the total bolometric luminosity of \sgra, $L_t
\sim 10^{36}$ erg/sec. In a single Compton scattering, the photon
energy on average increases by a factor $\gamma^2$, where $\gamma$ is
the Lorenz $\gamma$-factor of the electron taking part in the
up-scattering. Averaged over a relativistic Maxwellian electron
distribution with temperature $T_e$, the average boost factor becomes
$16 (k_B T_e/m_e c^2)^2$, where $k_B$ is the Boltzmann constant, and
$m_e$ is the electron mass. The typical frequency of a photon emitted
by the star is around $\nu = 3 k_B T_*/h \approx 2 \times 10^{15}$ Hz,
where $T_* = 30,000$ Kelvin is the stellar temperature, and $h$ is the
Planck constant. {\em Chandra} frequency band is centered around $\nu
\sim 10^{18}$ Hz.  Therefore electrons with temperature of $T_e
\approx 3\times 10^{10} K$ are the most efficient in scattering of S2
radiation into the {\em Chandra} band.

Such high temperature electrons are present in both the canonical ADAF
(without winds) model, e.g. \cite{Narayan02}, and even more so in the
less dense but hotter Non-Radiative Accretion Flows (NRAFs) that
incorporate effects of gas outflows. Since \sgra\ is very dim in
X-rays in its normal ``quiescent'' state, e.g. $L_X \sim
10^{33}$~erg~s$^{-1}$, there is then a hope of detecting a comptonized
``echo'' of a close star passage.

\section{Analytical estimates}\label{sec:estimates}

 We shall rely on the accretion flow models by \cite{Narayan02} and
\cite{Yuan03}, who used $M_{\rm BH} = 2.6 \times 10^6\msun$. Recently,
\sgra\ mass has been determined to be larger, e.g. $M_{\rm BH} =
(3-4)\times 10^6\msun$ \citep{Genzel03,Ghez03a}. We thus use an
intermediate value of $M_{\rm BH} = 3 \times 10^6\msun$.

The Thomson optical depth, $\tau_T$, of the hot accretion flow in
\sgra\ is quite small: $\tau_T \sim 10^{-3} n_8 r_1$, where $n_8 =
n_e/10^8$ is the typical electron density in units of
$10^8$~cm$^{-3}$, and $r_1 = R/10 R_S$. (Note that further out the
optical depth of the flow is even smaller.) Therefore, only the first
order scattering of the stellar photons need be considered.

At the moment of the closest approach to the SMBH, the star S2 in
particular, was $D = R_p \sim 2000 R_S$ away from \sgra, where $R_S =
2GM_{\rm BH}/c^2$ is the Schwarzschild radius of the SMBH, $D$ is the
current distance between the star and \sgra, and $R_p$ is the
pericenter distance of the orbit. The radiation energy density near
the SMBH is thus
\begin{equation}
u_{\rm rad} = \frac{L_*}{4 \pi D^2 c}\approx 10^{-3}
\frac{L_5}{D_{2000}^2}\;\hbox{erg sec}^{-1}\;,
\label{us2}
\end{equation}
where $L_5 = L_*/10^5 L_\odot$, $D_{2000} = D/(2000 R_S)$.

We now introduce $N_{\rm he}$ as the total number of electrons in the
hot accretion flow that are able to up-scatter an optical/UV photon
with frequency $\nu_* \simeq 2 \times 10^{15}$ Hz into the {\em
Chandra} band, $\nu_x \simeq 10^{18}$ Hz. The corresponding mass of
the accretion flow ions, assumed to be all protons for simplicity, is
clearly $\tilde M_{acc} \approx m_p N_{\rm he}$, where $m_p$ is the
proton mass.  Since scattering conserves number of photons, we can
estimate the X-ray luminosity of the up-comptonized S2 emission as

\begin{equation}
L_x \sim N_{\rm he} u_{\rm rad} c \sigma_T \frac{\nu_x}{\nu_{*}}\;
\approx 10^{33} \; \frac{M_{-10} L_5}{D_{2000}^2}\;\hbox{erg sec}^{-1}\;,
\end{equation}
where $M_{-10}\equiv \tilde M_{acc}/10^{-10}\msun$. Recall that
quiescent X-ray luminosity of \sgra\ is $L_x \simeq 2 \times 10^{33}$
erg/sec. Thus, if the region of accretion flow with $T_e \sim 3\times
10^{10}K$ had a mass of around $\tilde M_{acc}= 10^{-10} \msun$ or
larger, the up-comptonized S2 emission would have been measurable in
2002 by {\em Chandra}.

We can relate the accretion flow mass $M_{acc}$ at a given
dimensionless radius $r$ to the accretion rate as 
\begin{equation}
M_{acc} \sim \dot M t_{\rm visc}\;,
\end{equation}
where $t_{\rm visc}$ is the viscous time estimated in
$\alpha$-viscosity accretion flows as
\begin{equation}
t_{\rm visc} \simeq t_{\rm dyn} \alpha^{-1} \frac{R^2}{H^2}\; \simeq
30 \;\hbox{sec}\; r^{3/2} \alpha^{-1} \frac{R^2}{H^2}\;,
\end{equation}
where $t_{\rm dyn} = R/v_K$ is the local dynamical time and $H$ is the
geometrical thickness of the accretion flow. For flows considered here
$H/R \sim 1$. For example, with an accretion rate of $\dot M = 10^{-7}
\dot M_{-7} \msun$~year$^{-1}$, we have
\begin{equation}
M_{acc} \simeq 3\times 10^{-11} \msun \dot M_{-7} \alpha_{-1}^{-1} r_1^{3/2}\;,
\end{equation}
where $\alpha_{-1}=\alpha/0.1$. Requiring $M_{acc} \simlt 10^{-10}
\msun$, we have
\begin{equation}
\dot M \simlt 3\times 10^{-7} \alpha_{-1} r_1^{3/2}\;.
\label{mdcrit}
\end{equation}
This is a realistic range of accretion rates for \sgra\
\citep{Narayan02,Yuan03}. 

\section{Calculation of the comptonized spectrum}\label{sec:spectrum} 

The estimates above are useful but nonetheless a more careful
calculation is required as an accretion flow spans a range of
temperatures and its density profile may be changing in a non-trivial
way due to, e.g., outflows. For definitiveness, we shall use the
accretion flow structure as available in models of \cite{Narayan02}
and \cite{Yuan03}.

The electron temperature is at most equal to the local virial
temperature, e.g. temperature at which the thermal energy of the gas
is equal to the gravitational potential energy: $T_{\rm vir} = (2/3)
GM_{\rm BH} m_p/(k_B R) = 3.6\times 10^{12} r^{-1}$ K. Thus, $T_e \le
4 \times 10^9$ K for $r \ge 10^3$. Accordingly, we do not consider
Compton scattering in the outer, $r \ge 10^3$, regions of the flow.
In practice $T_e$ is still smaller than $T_{\rm vir}$, and hence it is
only the inner few hundred \schw radii of the flow that are important.
Since these radii are much smaller than S2 pericenter distance, $R_p \sim
2000 R_S$, we can safely assume that $u_{\rm rad} = $ const in this
optically thin region.

The emitted comptonized spectrum is calculated as
\begin{equation}
L_{\nu} = c \int_{R_{\rm i}}^{R_{\rm out}} d V n_e(R)
\int_{0}^{\infty} d \nu_1 \frac{d u_{\rm rad}(\nu_1)}{d \nu_1}
\sigma(\nu, \nu_1, T_e)\;,
\end{equation}
where $d V = 4\pi R^2 dR$ is the volume element, $\nu_1$ is the
frequency of the photon before scattering and $\nu$ is that after the
scattering, and $\sigma(\nu, \nu_1, T_e)$ is the cross section for
such a scattering averaged over the electron distribution function
with temperature $T_e$. The cross section is taken from
\cite{Nagirner}. We assume also that the bolometric luminosity of S2
is $L_* = 10^5 \lsun$ and the spectrum is that of a blackbody at
temperature $T = 30,000$ Kelvin.

\begin{figure}
\centerline{\psfig{file=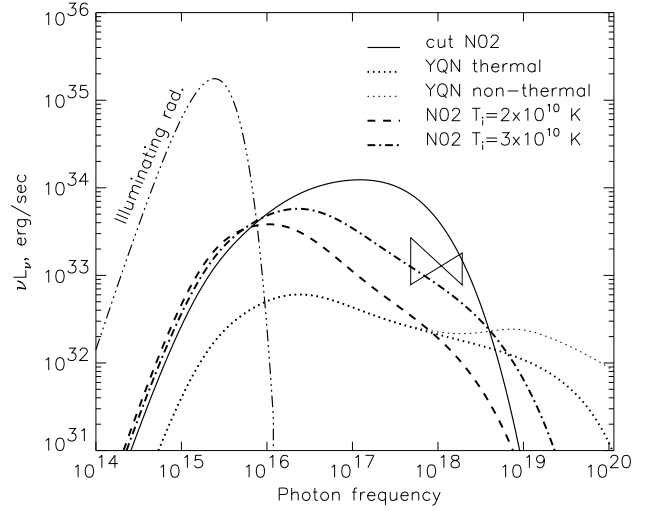,width=.49\textwidth}}
\caption{The up-comptonized emission of the close star S2 during its
closest approach to \sgra\ in 2002 for five different accretion flows,
as marked on the Figure. The spectrum of the star illuminating the
inner 100 \schw radii of the flow is also shown for comparison. The
{\em Chandra} data with uncertainties \citep{Baganoff03a} are shown
with the bow-tie.}
\label{fig:fig1}
\end{figure}

Figure \ref{fig:fig1} shows several examples of the expected
up-comptonized spectrum calculated for several different accretion
flow models. The dash-multiple-dot curve labelled by ``Illuminating
rad.'' is the stellar radiation incident on the inner $100 R_S$
sphere, plotted as a reference scale. The bow-tie is the {\em Chandra}
data on emission of \sgra\ with the spectral uncertainty
\citep{Baganoff03a}.

The solid curve is the simplest ``cut-off'' ADAF model used by
\cite{Narayan02} to illustrate the need for a two-temperature nature
of the accretion flow in \sgra. In this model the electron temperature
at large radii follows the virial gas temperature but is capped at
$T_e = 10^{10}$ K at smaller radii. We assumed $\alpha_{-1} = 1$ and
accretion rate of $\dot M = 3 \times 10^{-6} \msun$~year$^{-1}$, the
lower of the values used by \cite{Narayan02} (see his equation 3). The
model clearly over-predicts the emission in the X-ray band. This
occurs despite the fact that electrons never reach the temperature of
$3\times 10^{10}$ K. The electrons in the tail of the Maxwellian
distribution produce the {\em Chandra} band X-rays, and since the
accretion rate is an order of magnitude higher than the critical one
estimated in equation \ref{mdcrit}, the model manages to emit a
measurable X-ray flux.

Next we test the more refined ADAF models by \cite{Narayan02}. These
employ a self-consistent description of the electron-ion Coulomb
energy exchange rather than the simple temperature cutoff. In
particular, we approximate the electron temperature profiles in his
Figure 5 as 
\begin{equation}
T_e = \cases{T_0 \left[r/3\right]^{-1/2},\hskip 0.5cm \hbox{if}\quad
T_e \; < \;
T_{\rm vir}(r)\cr
T_{\rm vir}(r), \hskip 1cm \hbox{otherwise}\;,\cr}
\end{equation}
where $T_0$ is the electron temperature at $r=3$. In particular, we
consider two values of $T_0$ that are approximately consistent with
Figure 5 of \cite{Narayan02}: $T_0 = 2\times 10^{10}$ and $T_0 =
3\times 10^{10}$. Both models are for $\dot M = 10^{-6}
\msun$~year$^{-1}$ and $\alpha=0.1$. The higher initial electron
temperature model is probably ruled out by the existing {\em Chandra}
data on \sgra\ that show no noticeable variation in the X-ray flux
during 2002 (Baganoff 2004, private communication). The lower
temperature model may be however below what could have been
detected by {\em Chandra}.

Finally, we also test the more recent and more physically complete
NRAF models for \sgra\ by \cite{Yuan03}. Their Figure 2 shows the
electron temperature and the density profiles for one particular set
of parameters that fit \sgra\ broad band spectrum quite well. Due to
strong thermally driven outflows, the radial profile of density is
$\rho(r) \propto r^{-3/2 + s}$, with $s = 0.27$. Note that for ADAFs
$s=0$. The temperature profile in \cite{Yuan03} is quite close to a
power-law with $T(r) \propto r^{-1 + s} \simeq r^{-3/4}$. The
accretion rate close to the SMBH in this model is $\dot M(3 R_S) = 4
\times 10^{-8} \msun$~year$^{-1}$. 

\cite{Yuan03} also include a non-thermal component in the electron
distribution. As we find that the latter does not produce a dominant
contribution to the up-comptonized S2 emission in the {\em Chandra}
band, and since we do not have access to the exact model by
\cite{Yuan03}, we set the minimum and maximum $\gamma$-factors as
following: $\gamma_{\rm min} = 30$ and $\gamma_{\rm max} = 2000$, in a
rough agreement with Figure 2 of \cite{Yuan03}.

Figure \ref{fig:fig1} shows the resulting comptonized spectrum (fine
dotted curves) at the pericenter of S2 orbit for the \cite{Yuan03}
model. The higher of the curves includes both thermal and non-thermal
electrons, whereas the other one includes scattering due to the
thermal electrons only. In both cases the resulting X-ray emission is
about a factor of 10 below the quiescent X-ray emission of \sgra, so
that the model of \cite{Yuan03} clearly passes the test. 

\section{Lightcurve of the event}

The trajectory of S2 is calculated based on the best fit to its orbit
by \cite{Ghez03a}. The background optical-UV radiation energy density
near \sgra\ location is assumed to be dominated by the whole of the
\sgra\ star cluster, whose luminosity is of order $L_{\rm cl} = 10^{7}
\lsun$, and whose size is $R_{\rm cl} \simeq 0.2$ parsec
\citep{Genzel03}. Thus, $u_{\rm bg} = L_{\rm cl}/4\pi R_{\rm cl}^2$. We
assume a spectrum identical to that of S2 for simplicity. The
background is quite negligible in comparison with the radiation energy
density $u_{\rm rad}$ of S2 during its pericenter passage (equation
\ref{us2}).

\begin{figure}
\centerline{\psfig{file=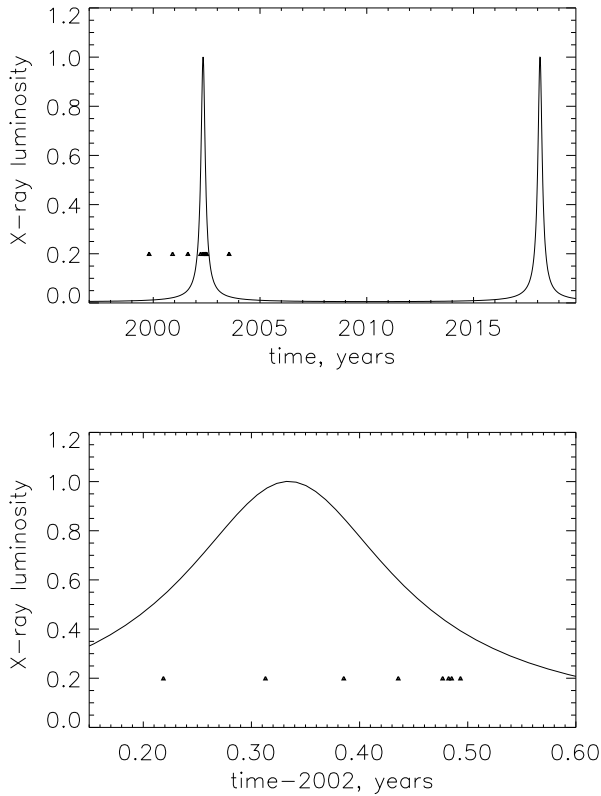,width=.49\textwidth}}
\caption{The lightcurve of the optical/UV radiation of the close star
S2 up-comptonized by the accretion flow into the X-ray band. The upper
panel shows the lightcurve on time scales of tens of years, while the
lower one shows the emission during 2002 when S2 passes through the
pericenter of its orbit. The filled triangles mark the epochs when
{\em Chandra} observed the Galactic Center region
\citep{Park04,Eckart04}. The y-coordinate of the triangles in the
Figure is arbitrary.}
\label{fig:fig2}
\end{figure}

Since the cooling time of the hot electrons due to inverse Compton
effect is very long compared to the local dynamical time
\citep{Chang03}, it can be neglected. The X-ray lightcurve is then
\begin{equation}
L_x = L_{x, p} \;\left[\frac{R_p}{D}\right]^2\;,
\end{equation}
where $D$ is the current distance between the SMBH and S2.  The
resulting X-ray lightcurve, normalized to the maximum reached at the
pericenter, $L_{x, p}$, is plotted in Figure \ref{fig:fig2}. The upper
panel shows the lightcurve on the scales of tens of years, whereas the
lower panel concentrates on that during the year 2002 only. Clearly
the duration of the event is several months. Very luckily, {\em
Chandra} observed \sgra\ (filled triangles in Figure \ref{fig:fig2})
on many occasions in 2002 (e.g. see \cite{Park04}). It is thus possible
that a careful analysis of {\em Chandra} 2002 data would be able to
constrain the amount of an extra up-comptonized X-ray flux to a small
fraction of the quiescent X-ray luminosity of \sgra. Such an analysis
may be facilitated by the fact that the lightcurve of the event is
well determined due to small overall errors in the trajectory of the
star S2 \citep{Schoedel02, Ghez03b}.

\section{Discussion}\label{sec:discussion}

We have shown that up-comptonization of the optical/UV radiation of
the star S2, which famously passed through the pericenter of its orbit
only $\sim 2000$ \schw radii away from \sgra\ \citep{Schoedel02},
constrains the properties of the accretion flow in the immediate
vicinity of the SMBH. These constraints are completely independent
from those due to the rotation measure limits in the radio wavelengths
\citep{Bower03}, and are thus complimentary to those results. We urge
the observers to search the {\em Chandra} 2002 data for signatures of
the transient brightening with the lightcurve shown in Figure
\ref{fig:fig2}. Such an analysis may perhaps yield much tighter
constraints on the amplitude of the effect than Figure \ref{fig:fig1}
indicates.

Future observations of close star passages may be more constraining
that that from the year 2002 if (i) the star approaching the SMBH is
brighter and/or closer to \sgra\ than $2000 R_S$; (ii) if more
sensitive and well planned X-ray observations of the close stellar
passage are made. In addition, interaction of stellar winds with the
surrounding hot accretion flow may yield observable signatures in a
range of wavelengths depending on the stellar mass loss rate and the
outflow velocity and (unfortunately) not well constrained details of
presumably collisionless wind/accretion flow shocks (e.g., see \cite{Loeb04}).

The idea of using the close star passages to constrain the properties
of the accretion flow around \sgra\ is not new. \cite{Chang03} noted
that the inverse Compton effect will provide an additional cooling of
the hot electrons. However the corresponding relative change in the
electron temperature is only $\simlt 10^{-4}$ and appears very hard to
observe in any wavelength. Here we showed that the up-comptonized S2
radiation itself, on the over hand, is much easier to observe since
\sgra\ quiescent X-ray luminosity is very small.  In addition,
\cite{NS03} and \cite{Cuadra03} have shown that passages of close
stars could also be used to constrain the properties of a cold
geometrically thin ``inactive'' accretion disk around \sgra. The fact
that no near infra-red ``echoes'' or stellar eclipses were observed up
to date probably indicates that such a disk does not presently exist
in \sgra.

Concluding, we note that close stellar passages are of
significant value for solving the riddle of \sgra\ and collisionless
gas accretion onto SMBHs in general through a variety of physical
processes.

\del{Density of the wind:
\begin{equation}
n_w \sim M_w/4\pi R^2 v_w m_p \sim 5\times 10^4 cm^{-3} \dot M_{-6}
r_3^{-2} v_8^{-1}
\end{equation}
Mass of the wind:
\begin{equation}
M \sim 10^{-6} \msun/year \times 0.1 year = 2 \times 10^{26} g
\end{equation}
Mass of the non-radiative flow $\sim 5 (10^{15})^3 \times 2 10^4
\times 2 10^{-24} = 2 \times 10^{26}$ g also.}

\acknowledgement The author acknowledges useful discussions with Fred
Baganoff and Henk Spruit.

\bibliographystyle{aa}

\begin{thebibliography}{20}
\expandafter\ifx\csname natexlab\endcsname\relax\def\natexlab#1{#1}\fi

\bibitem[{{Baganoff} {et~al.}(2001){Baganoff}, {Bautz}, {Brandt}, {Chartas},
  {Feigelson}, {Garmire}, {Maeda}, {Morris}, {Ricker}, {Townsley}, \&
  {Walter}}]{Baganoff01}
{Baganoff}, F.~K., {Bautz}, M.~W., {Brandt}, W.~N., {et~al.} 2001, \nat, 413,
  45

\bibitem[{{Baganoff} {et~al.}(2003){Baganoff}, {Maeda}, {Morris}, {Bautz},
  {Brandt}, {Cui}, {Doty}, {Feigelson}, {Garmire}, {Pravdo}, {Ricker}, \&
  {Townsley}}]{Baganoff03a}
{Baganoff}, F.~K., {Maeda}, Y., {Morris}, M., {et~al.} 2003, \apj, 591, 891

\bibitem[{{Bower} {et~al.}(2003){Bower}, {Wright}, {Falcke}, \&
  {Backer}}]{Bower03}
{Bower}, G.~C., {Wright}, M.~C.~H., {Falcke}, H., \& {Backer}, D.~C. 2003,
  \apj, 588, 331

\bibitem[{{Chang} \& {Choi}(2003)}]{Chang03}
{Chang}, H. \& {Choi}, C. 2003, \aap, 410, 519

\bibitem[{{Cuadra} {et~al.}(2003){Cuadra}, {Nayakshin}, \&
  {Sunyaev}}]{Cuadra03}
{Cuadra}, J., {Nayakshin}, S., \& {Sunyaev}, R. 2003, \aap, 411, 405

\bibitem[{{Eckart} {et~al.}(2004){Eckart}, {Baganoff}, {Morris}, {Bautz},
  {Brandt}, {Garmire}, {Genzel}, {Ott}, {Ricker}, {Straubmeier}, {Viehmann}, \&
  {Sch{\" o}del}}]{Eckart04}
{Eckart}, A., {Baganoff}, F.~K., {Morris}, M., {et~al.} 2004, ArXiv
  Astrophysics e-prints

\bibitem[{{Genzel} {et~al.}(2003){Genzel}, {Sch{\" o}del}, {Ott}, {Eisenhauer},
  {Hofmann}, {Lehnert}, {Eckart}, {Alexander}, {Sternberg}, {Lenzen}, {Cl{\'
  e}net}, {Lacombe}, {Rouan}, {Renzini}, \& {Tacconi-Garman}}]{Genzel03}
{Genzel}, R., {Sch{\" o}del}, R., {Ott}, T., {et~al.} 2003, \apj, 594, 812

\bibitem[{{Ghez} {et~al.}(2003{\natexlab{a}}){Ghez}, {Becklin}, {Duch{\^ e}ne},
  {Hornstein}, {Morris}, {Salim}, \& {Tanner}}]{Ghez03a}
{Ghez}, A.~M., {Becklin}, E., {Duch{\^ e}ne}, G., {et~al.} 2003{\natexlab{a}},
  to be published in Astron. Nachr., Vol. 324, No. S1 (2003), Special
  Supplement ``The central 300 parsecs of the Milky Way'', Eds. A. Cotera, H.
  Falcke, T. R. Geballe, S. Markoff, (astro-ph/0303151)

\bibitem[{{Ghez} {et~al.}(2003{\natexlab{b}}){Ghez}, {Duch{\^ e}ne},
  {Matthews}, {Hornstein}, {Tanner}, {Larkin}, {Morris}, {Becklin}, {Salim},
  {Kremenek}, {Thompson}, {Soifer}, {Neugebauer}, \& {McLean}}]{Ghez03b}
{Ghez}, A.~M., {Duch{\^ e}ne}, G., {Matthews}, K., {et~al.} 2003{\natexlab{b}},
  \apj, 586, L127

\bibitem[{{Loeb}(2004)}]{Loeb04}
{Loeb}, A. 2004, \mnras, 350, 725

\bibitem[{{Nagirner} \& {Poutanen}(1994)}]{Nagirner}
{Nagirner}, D.~I. \& {Poutanen}, J. 1994, {Single Compton scattering}
  (Amsterdam: Harwood Academic Publishers, |c1994)

\bibitem[{{Narayan}(2002)}]{Narayan02}
{Narayan}, R. 2002, in Lighthouses of the universe: the most luminous celestial
  objects and their use for cosmology, ed. M.Gilfanov, R. Sunyaev \& E.
  Churazov (Berlin: Springer), 405

\bibitem[{{Nayakshin} \& {Sunyaev}(2003)}]{NS03}
{Nayakshin}, S. \& {Sunyaev}, R. 2003, \mnras, 343, L15

\bibitem[{{Park} {et~al.}(2004){Park}, {Muno}, {Baganoff}, {Maeda}, {Morris},
  {Howard}, {Bautz}, \& {Garmire}}]{Park04}
{Park}, S., {Muno}, M.~P., {Baganoff}, F.~K., {et~al.} 2004, \apj, 603, 548

\bibitem[{{Quataert}(2002)}]{Quataert02}
{Quataert}, E. 2002, \apj, 575, 855

\bibitem[{{Reid} {et~al.}(1999){Reid}, {Readhead}, {Vermeulen}, \&
  {Treuhaft}}]{Reid99}
{Reid}, M.~J., {Readhead}, A.~C.~S., {Vermeulen}, R.~C., \& {Treuhaft}, R.~N.
  1999, \apj, 524, 816

\bibitem[{{Ruszkowski} \& {Begelman}(2002)}]{Ruszkowski02}
{Ruszkowski}, M. \& {Begelman}, M.~C. 2002, \apj, 573, 485

\bibitem[{{Sch{\" o}del} {et~al.}(2002){Sch{\" o}del}, {Ott}, {Genzel},
  {Hofmann}, {Lehnert}, {Eckart}, {Mouawad}, {Alexander}, {Reid}, {Lenzen},
  {Hartung}, {Lacombe}, {Rouan}, {Gendron}, {Rousset}, {Lagrange}, {Brandner},
  {Ageorges}, {Lidman}, {Moorwood}, {Spyromilio}, {Hubin}, \&
  {Menten}}]{Schoedel02}
{Sch{\" o}del}, R., {Ott}, T., {Genzel}, R., {et~al.} 2002, \nat, 419, 694

\bibitem[{{Yuan} {et~al.}(2002){Yuan}, {Markoff}, \& {Falcke}}]{Yuan02}
{Yuan}, F., {Markoff}, S., \& {Falcke}, H. 2002, \aap, 383, 854

\bibitem[{{Yuan} {et~al.}(2003){Yuan}, {Quataert}, \& {Narayan}}]{Yuan03}
{Yuan}, F., {Quataert}, E., \& {Narayan}, R. 2003, \apj, 598, 301

\end{thebibliography}

\end{document}